\documentclass[conference]{IEEEtran}
\usepackage[hyphens]{url}
\usepackage{hyperref}
\hypersetup{breaklinks=true,colorlinks,allcolors=blue}
\usepackage{doi}
\usepackage{orcidlink}
\usepackage{balance}
\usepackage{cite}

\IEEEoverridecommandlockouts
\usepackage{array}

\begin{document}

\title{Toward Secure and Compliant AI: Organizational Standards and Protocols for NLP Model Lifecycle Management%
\thanks{This is the author-accepted manuscript of a paper accepted to ICAART 2026.
The final published version will appear in the conference proceedings.}
}

\author{\IEEEauthorblockN{Sunil Arora\,\orcidlink{0009-0007-3066-3461}, John Hastings\,\orcidlink{0000-0003-0871-3622}}
  \IEEEauthorblockA{
    \textit{The Beacom College of Computer \& Cyber Sciences} \\
    \textit{Dakota State University, USA}\\
    sunil.arora@trojans.dsu.edu, john.hastings@dsu.edu \\
    }
}

\maketitle

\begin{abstract}
 Natural Language Processing (NLP) systems are increasingly used in sensitive domains such as healthcare, finance, and government, where they handle large volumes of personal and regulated data. However, these systems introduce distinct risks related to security, privacy, and regulatory compliance that are not fully addressed by existing AI governance frameworks. This paper introduces the Secure and Compliant NLP Lifecycle Management Framework (SC-NLP-LMF)\textbf{,} a comprehensive six-phase model designed to ensure the secure operation of NLP systems from development to retirement. The framework, developed through a systematic PRISMA-based review of 45 peer-reviewed and regulatory sources, aligns with leading standards, including NIST AI RMF, ISO/IEC 42001:2023, the EU AI Act, and MITRE ATLAS. It integrates established methods for bias detection, privacy protection (differential privacy, federated learning), secure deployment, explainability, and secure model decommissioning. A healthcare case study illustrates how SC-NLP-LMF detects emerging terminology drift (e.g., COVID-related language) and guides compliant model updates. The framework offers organizations a practical, lifecycle-wide structure for developing, deploying, and maintaining secure and accountable NLP systems in high-risk environments.
\end{abstract}


\section{\uppercase{Introduction}}
\label{sec:introduction}

The exponential proliferation of natural language processing (NLP) technologies across critical industries such as healthcare, finance, law, and public administration has intensified the need for comprehensive management of NLP models throughout their operational lifecycles. As these models increasingly influence decision-making processes, especially in high-risk environments with direct societal and legal implications \cite{Malek2022}, ensuring their security, compliance, transparency, and robustness has become a paramount organizational responsibility \cite{Eu2024AIAct,veale2021eu}. 
For example, in July 2025, a significant AI security vulnerability, now cataloged as CVE-2025-8217, was identified in Amazon Q Developer for Visual Studio Code, a widely used AI-powered coding assistant \cite{aws2025_q_incident,mcnamara2025_AWSq}. The breach was enabled by an overly permissive GitHub token embedded in AWS CodeBuild configurations, which allowed an attacker to commit malicious prompt-based code into the extension's open-source repository. The injected payload was designed to manipulate the system via natural language instructions, allowing it to wipe file systems and delete cloud infrastructure. Although a syntax error prevented its execution, the compromised version (v1.84.0) was distributed through the Visual Studio Code Marketplace, potentially affecting up to one million developers worldwide. This incident exemplifies a growing class of “prompt injection” attacks targeting the semantic interpretation layers of NLP-driven AI tools. Unlike traditional code injection exploits, these attacks leverage indirect linguistic manipulation to subvert model behavior and initiate high-impact operations through natural language prompts. The Amazon Q breach highlights the crucial need for secure development pipelines, model-level prompt validation, and adversarial testing in NLP system development lifecycles. This is not an isolated case, and there are many notable and significant incidents. Earlier, in 2016, Microsoft’s Tay chatbot became a cautionary example of inadequate NLP safety measures. Within hours of public interaction, adversarial users manipulated Tay’s conversational logic, prompting it to generate racist and offensive outputs, and it was shut down within 24 hours \cite{neff,Perez2016Tay}. The incident underscored the risks of exposing unfiltered NLP systems to open-domain inputs without effective content moderation, adversarial defense, or behavioral monitoring. 

There are multiple general AI governance frameworks \cite{airmf,iso2020,ENISA_2023multilayer,UKAICyberCode2025} that have proposed foundational standards for trustworthy AI systems. However, these frameworks often lack the necessary granularity and specificity to address the distinct challenges posed by NLP models, including linguistic drift, contextual bias evolution, and inadvertent privacy breaches.

A continuous governance and management framework is required to support the dynamic nature of NLP systems. This framework must be dynamic and able to manage the adoption of new risks, evolve in response to changing compliance requirements, and adapt to shifting operational contexts. Organizations navigating increasingly complex regulatory landscapes, including the EU AI Act \cite{europeancommission2021}, which classifies many NLP applications as ``high-risk" systems, require a structured, proactive, and domain-specific approach to NLP lifecycle management.

This paper proposes the Secure and Compliant NLP Lifecycle Management Framework (SC-NLP-LMF), a comprehensive, multi-phase protocol for governing NLP models throughout their entire lifecycle. Building upon best practices in AI governance, bias auditing, privacy-preserving techniques, and dynamic model monitoring, the SC-NLP-LMF provides actionable guidance for organizations seeking to operationalize trustworthy NLP deployment at scale. The proposed framework addresses the limitations of existing approaches by systematically integrating governance structures, technical controls, monitoring, and compliance mechanisms. It offers a clear path toward secure, compliant, and resilient NLP systems.

\section{\uppercase{Related Work}}

The management of AI models throughout their operational lifecycles has become an increasingly prominent research topic, particularly in response to the complex societal, ethical, and legal implications of deploying ML and NLP technologies at scale. Although several foundational frameworks have been proposed for general AI system governance, they often fall short in addressing the unique, dynamic risks associated with NLP systems, such as semantic drift, contextual bias evolution, and privacy vulnerabilities. 

General AI lifecycle and security standards have primarily focused on high-level governance principles. The National Institute of Standards and Technology’s AI Risk Management Framework \cite{airmf} provides a structured approach to managing AI risks through governance, mapping, measurement, and management functions. Similarly, ISO/IEC 42001:2023 \cite{iso_42001_2023} provides a structured approach for AI governance at the organizational level, introducing the concept of an Artificial Intelligence Management System (AIMS) and ISO/IEC TR 24028:2020 \cite{iso2020} outlines key trustworthiness attributes for AI systems, including robustness, security, and explainability. Similarly, ENISA (European Union Agency for Cybersecurity) Multilayer Framework for Good Cybersecurity Practices for AI \cite{ENISA_2023multilayer} has outlined general cybersecurity best practices for AI systems. While these frameworks have established the groundwork for AI governance, they address AI systems at an abstract and high level without delving into domain-specific issues pertinent to NLP, such as language evolution and cultural variability. Furthermore, existing AI lifecycle management pipelines, such as TensorFlow Extended (TFX) \cite{baylor} and Kubeflow Pipelines \cite{Amershi}, primarily emphasize scalability and reproducibility in engineering workflows without integrating lifecycle compliance auditing or dynamic bias monitoring mechanisms. They do not provide a comprehensive lifecycle coverage designed to maintain an NLP model's compliance and operational security over time. 

Table \ref{tab:1} provides a comparative overview of existing AI lifecycle and governance frameworks and tools, highlighting their respective scopes, supported phases, available toolkits, and adoption challenges. As shown, none of the frameworks comprehensively address the full NLP lifecycle, particularly in regulated environments, underscoring the need for the proposed SC-NLP-LMF model. In light of the gaps identified in the current body of literature, including limited NLP-specific lifecycle frameworks, inadequate integration of security and compliance protocols, and insufficient operational support for phases such as decommissioning, retraining, and bias monitoring. This work directly addresses these deficiencies. The proposed SC-NLP-LMF builds upon and extends prior efforts by offering a lifecycle-wide, secure, and compliant management structure tailored specifically to NLP systems. In the following sections, we outline the methodology used to synthesize the framework, followed by its detailed architecture and evaluation.

\begin{table*}
\caption{Comparison of Major AI Lifecycle and Governance Frameworks}
    \label{tab:1} \centering
    \small
\begin{tabular}{|>{\centering\arraybackslash}p{0.14\linewidth}|>{\centering\arraybackslash}p{0.14\linewidth}|>{\centering\arraybackslash}p{0.19\linewidth}|>{\centering\arraybackslash}p{0.14\linewidth}|>{\centering\arraybackslash}p{0.19\linewidth}|}\hline
\textbf{Framework} & \textbf{Scope} & \textbf{Lifecycle Phases Covered} & \textbf{Practical Toolkits} & \textbf{Adoption Barriers} \\\hline

NIST AI RMF  \cite{airmf}& General AI & Risk planning, design, deployment, monitoring, governance & RMF Playbook, AI RMF Explorer, NIST Taxonomies & Limited NLP specificity; high institutional learning curve \\\hline

EU AI Act \cite{Eu2024AIAct}& General AI (Regulatory) & Design, risk classification, documentation, registration, market release & Compliance checklists, conformity assessment & Legal ambiguity; NLP-specific risks not yet standardized \\\hline

MITRE ATLAS \cite{mitre_atlas}& General AI & Threat modeling, attack mitigation  & Red Teaming, Adversary tactics, techniques, and case studies for artificial intelligence (AI) systems & Focused on threats; not a full governance model \\\hline

SMACTR \cite{raji}& General AI & Auditing, internal accountability, documentation & Role assignment maps, audit pipeline strategies & Organizational role clarity and maturity required \\\hline

ISO/IEC 42001:2023 \cite{iso_42001_2023}& General AI & Policy governance, data traceability, auditability & Certification schemas & Expensive for SMEs; high-level compliance focus \\\hline

ISO/IEC TR 24028:2020 \cite{iso2020}& General AI & Risk management, security, dependability & Trustworthiness of AI & Broad technical orientation; minimal implementation guidance \\\hline

Fairlearn \cite{bird2020fairlearn}& NLP and Structured Data & Evaluation, fairness auditing, trade-off analysis & Python library, interactive dashboard & Limited LLM support; pre-deployment focus \\\hline

Meta Frontier AI \cite{meta_frontier_2025}& General AI & Threat modeling, release controls, monitoring & Tiered governance layers, uplift trackers, red-teaming tools & Internal to Meta; partially open; early-stage development \\\hline

NCSC Secure AI Guidelines \cite{UKGov2025AICyberSecurity}& General AI & Secure design, development, deployment, monitoring & Threat modeling templates, deployment security checklists & Cybersecurity focus; minimal fairness or NLP lifecycle integration \\\hline

SC-NLP-LMF (this paper)& NLP-Specific & Full lifecycle: planning, training, deployment, monitoring, retraining, decommissioning & Integrated with Fairlearn, Audit Trails, secure deployment& Integration with legacy pipelines; organizational compliance capacity needed \\ \hline

\end{tabular}
 
\end{table*}

\section{\uppercase{Methodology}}

Building on the gaps highlighted in the prior section, particularly the lack of NLP-specific lifecycle standards, limited guidance on secure deployment, monitoring, and decommissioning protocols, this work introduces a unified framework to address these needs. The SC-NLP-LMF was developed through a rigorous multi-phase methodology designed to synthesize literature, align with existing standards, and ensure practical alignment with real-world NLP deployment and governance challenges.

The design of the SC-NLP-LMF adheres to a rigorous methodology, grounded in systematic literature analysis, critical synthesis of industry best practices, and normative alignment with emerging regulatory requirements. Recognizing the unique operational and compliance challenges posed by NLP systems, the methodology aimed to construct a domain-specific, actionable framework that extends beyond existing generalized AI governance structures.

The initial phase involved a systematic review of peer-reviewed literature from leading AI, NLP, and ML venues following Preferred Reporting Items for Systematic Reviews and Meta-Analyses (PRISMA) guidelines. PRISMA is a reporting guideline for systematic reviews and meta-analyses. Its primary aim is to improve the transparency and completeness of reporting, which in turn enhances the quality and reproducibility of these types of research \cite{Moher2009PRISMA}. Articles were selected based on relevance to model lifecycle management, bias auditing, adversarial robustness, data governance, and privacy-preserving techniques. Industry standards and government-issued regulatory documents, such as the NIST AI Risk Management Framework and the EU AI Act, were incorporated to ensure the proposed framework would align with imminent legal and compliance requirements. A structured review, conducted using the PRISMA methodology, retrieved 186 candidate documents from academic databases, standards bodies, and technical reports.

After applying inclusion and exclusion criteria, 45 high-quality documents were selected for final synthesis. 
We utilized Google Scholar, IEEE Xplore, ACM Digital Library, SpringerLink, industry, regulatory, and government websites to gather the white paper and scholarly articles. These span across foundational AI risk models, empirical fairness studies, AI frameworks, lifecycle protocols, and regulatory texts. To ensure relevance, rigor, and reproducibility, a structured set of inclusion and exclusion criteria was applied during the literature selection process. Included articles were required to be written in English and to present substantive contributions related to AI or NLP model lifecycle governance. Specifically, studies were retained if they proposed or evaluated AI lifecycle frameworks, described security or compliance tools for NLP models, AI or NLP security incidents, or documented real-world organizational practices related to AI audits, monitoring, or decommissioning. Only peer-reviewed journal papers, accepted conference proceedings, official technical white papers from recognized standards organizations or governmental agencies (e.g., ISO, NIST, ENISA), or technical papers for tools, protocols, or security incidents were included. This filtering process ensured that the resulting evidence base was both methodologically robust and directly aligned with the research objective of designing a secure and compliant NLP lifecycle framework. 

Building upon this foundation, the SC-NLP-LMF was synthesized according to five normative design principles: (1) Security-by-Design, integrating robustness against adversarial threats and data leakage; (2) Bias Auditing and Mitigation, requiring proactive bias measurement and remediation throughout the model lifecycle; (3) Privacy Preservation, ensuring that models are trained, validated, and deployed in adherence to privacy-protection standards such as GDPR; (4) Continuous Monitoring and Drift Detection, mandating dynamic performance validation and semantic drift audits post-deployment; and (5) Lifecycle Governance, emphasizing transparent documentation, traceability, and organizational accountability structures. These principles were chosen not only for their empirical support in the literature but also for their alignment with emerging legal compliance regimes.

The methodological construction also incorporated a critical evaluation of existing lifecycle tool chains and their limitations. For example, while TensorFlow Extended \cite{baylor} and Kubeflow Pipelines \cite{Amershi} provide robust model engineering workflows, they lack embedded bias auditing, compliance monitoring, and semantic drift management modules necessary for compliant NLP system operations. Similarly, although documentation practices such as Model Cards \cite{mitchell2019model} enhance transparency, they are static artifacts and do not address ongoing lifecycle risks or dynamic regulatory obligations.

\section{\uppercase{Secure and Compliant NLP Lifecycle Management  Framework (SC-NLP-LMF)}}

The SC-NLP-LMF introduced in this paper addresses the critical need for a structured, domain-specific protocol that guides the full lifecycle of NLP systems from data collection through model decommissioning while ensuring security, regulatory compliance, and ethical robustness. Unlike general-purpose AI lifecycle models that abstract away the particular risks associated with language-based systems, the SC-NLP-LMF integrates governance protocols, bias auditing, privacy-preserving mechanisms, and continuous evaluation processes specifically calibrated for NLP's dynamic nature and sociolinguistic impact.

At its core, the SC-NLP-LMF is composed of six interconnected lifecycle phases:  

\begin{itemize}
    \item \textbf{Data Governance}
    \item \textbf{Secure Model Training}
    \item \textbf{Deployment Governance}
    \item \textbf{Monitoring and Drift Detection}
    \item \textbf{Retraining and Updates}
    \item \textbf{Decommissioning and Archival}
\end{itemize}

\begin{figure}
    \centering
    \includegraphics[width=1\linewidth]{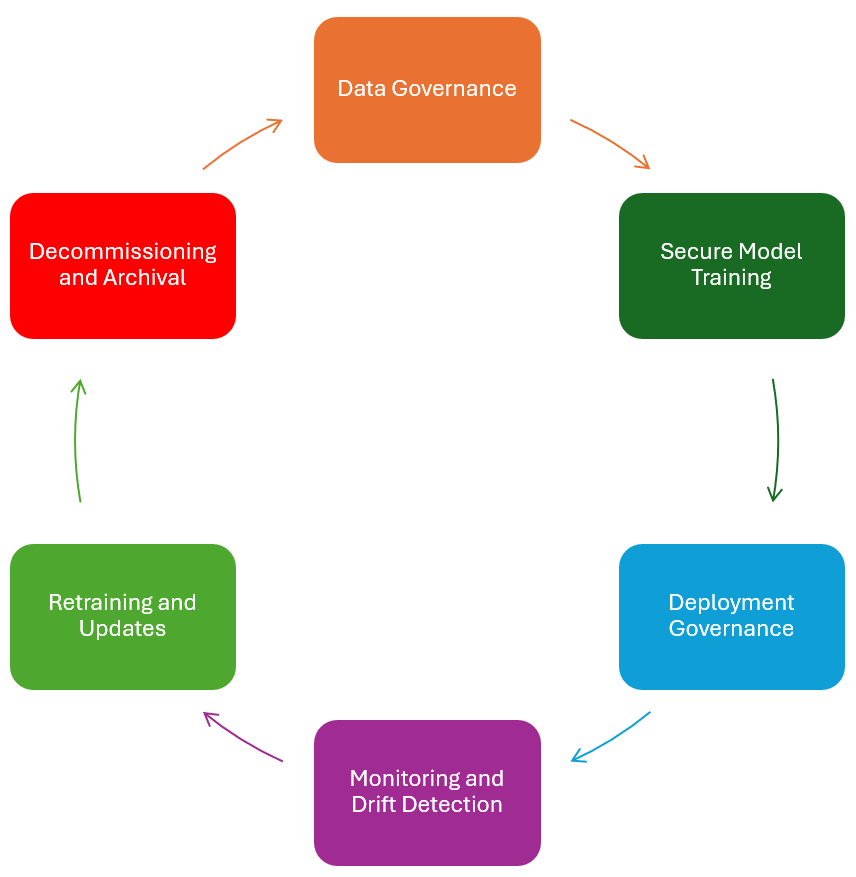}
    \caption{Six Phases of Secure and Compliant NLP Lifecycle Management (SC-NLP-LMF) Framework}
    \label{fig:1}
\end{figure}

Each phase includes concrete practices supported by academic research and emerging regulatory standards. Figure \ref{fig:1} provides a high-level schematic representation of the lifecycle, while Table \ref{tab:2} (Protocols and Standards Across the NLP Model Lifecycle) outlines specific tools and practices associated with each phase.

\begin{table*}[ht]
    \centering
    \caption{SC-NLP-LMF phases, associated standards and protocols}\label{tab:2} \centering
    \small
    \begin{tabular}{|>{\centering\arraybackslash}p{0.2\linewidth}|>{\centering\arraybackslash}p{0.4\linewidth}|>{\centering\arraybackslash}p{0.3\linewidth}|}\hline
         \textbf{Lifecycle Phase}	& \textbf{Key Protocols and Standards}	& \textbf{Supporting Frameworks and References}\\
\hline
Data Governance		& Data Statements for NLP  \cite{bender}, Differential Privacy \cite{trung,Martin}& EU AI Act \cite{Eu2024AIAct}\\\hline
Secure Model Training		& Fairness Audits (Fairness Indicators) \cite{xu2019fairness}, SHAP \cite{mosca-etal-2022-shap}, LIME \cite{feras,gitanjali}& SafeML \cite{Meghdad}\\\hline
Deployment Governance   & Model Cards for Model Reporting \cite{mitchell2019model}, Secure Deployment Practices \cite{google_saif}, Ethics guidelines for trustworthy AI \cite{eu-ethic-guideline}& Responsible AI Practices \cite{google_ai_principles,microsoftairesponsible}, Trustworthy AI Development \cite{brundage2020}\\\hline
Monitoring and Drift Detection   & Test, Evaluation, Verification, and Validation (TEVV) \cite{airmf}& NIST AI RMF \cite{airmf}, MITRE ATLAS \cite{mitre_atlas},  SMACTR Internal Audit Framework  \cite{raji}\\\hline
Retraining and Updates   & Bias Correction and Model Updates& AI TRiSM \cite{HABBAL2024122442}\\\hline
Decommissioning and Archival   & Secure Data and Model Disposal \cite{UKAICyberCode2025}& Verifiable Claims \cite{brundage2020}\\ \hline
    \end{tabular}
\end{table*}

The \textbf{Data Governance} phase anchors the framework by ensuring transparency, fairness, and legality at the point of data acquisition. This phase draws on Bender and Friedman \cite{bender} to mandate explicit documentation of dataset properties, including linguistic coverage, speaker demographics, and collection conditions. Privacy considerations are incorporated through differential privacy protocols \cite{McMahan}, and version-controlled data storage. Differential privacy is a framework for releasing statistical information about a dataset while ensuring the provable protection of individual privacy within that dataset. It achieves this by adding carefully calibrated noise to the data or query results, ensuring that the output of the analysis is statistically similar regardless of whether a single individual's data is included. This mechanism limits what can be inferred about any specific individual, regardless of the amount of other information known about them or the dataset \cite{trung,Martin}. These measures align closely with GDPR mandates regarding lawful and transparent data processing \cite{Eu2024AIAct}. 

The \textbf{Secure Model Training} phase operationalizes fairness and security as primary design objectives. Models must undergo bias audits using frameworks such as Fairness Indicators \cite{xu2019fairness} or IBM's AI Fairness 360 Toolkit \cite{bellamy}. Simultaneously, adversarial robustness testing is implemented to evaluate model resilience under perturbation \cite{papernot}. Similarly, SafeML is a framework that addresses privacy issues during model training \cite{Meghdad}. SafeML secures computations during model training using techniques like secret sharing and data masking, which protect sensitive information. It breaks down complex model training tasks into simple arithmetic operations, such as multiplication and addition, that can be performed over encrypted or hidden data. To ensure accuracy and resilience, SafeML introduces redundancy by assigning duplicate computation tasks to multiple nodes within a group. These nodes mask their results and share them with others for cross-verification. To enhance model transparency, SHAP (SHapley Additive exPlanations) and LIME (Local Interpretable Model-agnostic Explanations) can be used to generate detailed explanations and identify patterns of misclassification or feature sensitivity \cite{mosca-etal-2022-shap,feras,gitanjali}.  Compliance checkpoints, including adherence to lawful processing under GDPR Articles 5 and 22, are embedded during training, particularly for models operating in regulated domains such as healthcare or finance \cite{Eu2024AIAct}. 

The \textbf{Deployment Governance} phase introduces formal documentation and security standards as preconditions for releasing NLP models into production environments. It includes the publication of Model Cards \cite{mitchell2019model}, which provide stakeholders with transparent information about the model's intended use, limitations, and ethical implications. Secure deployment practices, including encrypted model serving, rate-limiting, and access controls, mitigate the risks of misuse or leakage. This phase also introduces ethical disclosures aligned with the AI FactSheets 360 initiative \cite{Arnold} and internal governance guidelines proposed by companies such as Google \cite{google_saif} and Microsoft \cite{microsoftairesponsible}. Meta’s Frontier AI Framework \cite{meta_frontier_2025} introduces a tiered governance model, risk thresholds, and outcomes-led risk evaluation process specifically tailored for high-capability models. In addition, the EU Ethics guidelines for trustworthy AI provide requirements for trustworthy AI deployment and a list of Trustworthy AI assessment criteria to help deploy and operationalize the key requirements \cite{eu-ethic-guideline}.  

Post-deployment, the \textbf{Monitoring and Drift Detection} phase institutes continuous risk assessment mechanisms. Drawing from research on semantic drift \cite{lazaridou2021mind}, models are continuously evaluated for performance degradation caused by language shifts, demographic changes, or evolving domain-specific terminology. Bias monitoring is reactivated at runtime, supported by periodic shadow evaluations and benchmarking against fairness baselines \cite{raji}. It is achievable through the SMACTR (Scoping, Mapping, Artifact Collection, Testing, and Reflection) framework, a structured, internal auditing protocol for AI accountability. SMACTR guides organizations through systematic evaluation across the AI development lifecycle. Unlike traditional ethics checklists, SMACTR enables ongoing, institutionalized review processes tailored to the sociotechnical realities of AI deployment. Its emphasis on internal roles, transparency, and impact assessment makes it particularly suited for the post-deployment monitoring and drift detection phases of secure NLP systems. Compliance is maintained through dynamic conformance checks mapped to the NIST AI Risk Management Framework \cite{airmf}, encouraging active measurement and risk mitigation as part of ongoing governance using AI TEVV (Test, Evaluation, Verification, and Validation). TEVV is a framework essential for building trustworthy systems, particularly AI and autonomous technologies, by ensuring they are reliable, safe, and fit for purpose throughout their lifecycle. Testing identifies failures, and evaluation assesses capabilities and risks in real-world scenarios. Verification confirms that the system is built correctly to meet requirements. Validation ensures that the right system is built to fulfill its intended uses. Applying TEVV helps mitigate risks, improve system quality, comply with standards, and ultimately enhance safety and trust in increasingly complex technological systems.

 Indicators from the monitoring layer trigger the \textbf{Retraining and Updates }phase, supporting the responsible adaptation of the model to new linguistic or operational environments. Key principles include dataset refresh with current data, re-execution of bias audits, and privacy-preserving learning procedures, such as those proposed in the federated learning paradigm \cite{McMahan}. Bias and fairness issues must be detected, mitigated, and tested in all phases of the NLP lifecycle. If detected, appropriate measures must be implemented to mitigate the issue \cite{fletcher2021}. Retraining must also account for user feedback loops and error reports, enabling iterative alignment with stakeholder expectations and regulatory shifts. Recent research emphasizes that retraining must go beyond performance optimization and embed robust mechanisms for bias mitigation and concept drift detection. In the articles \cite{Ray_2025} and \cite{HABBAL2024122442}, authors highlight the need for unified AI Trust, Risk, and Security Management (AI TRiSM) frameworks to govern retraining triggers, validate update justifications, and flag residual harm risks. AI TRiSM stands for Artificial Intelligence Trust, Risk, and Security Management. It is a framework developed by Gartner \cite{litan2024trust} to ensure AI model governance, trustworthiness, fairness, reliability, robustness, efficacy, and data protection.

Finally, the final phase of \textbf{Decommissioning and Archiva}l outlines a secure and responsible method for removing and archiving outdated, compromised, or no longer needed models. Audit logs must be preserved for retrospective analysis, and structured decommissioning protocols should be followed to ensure model artifacts do not persist in production systems. This is consistent with organizational AI lifecycle closure and claim-verifiability guidance recommended in \cite{brundage2020}. This is important, particularly in relation to post-deployment audit trails, data governance, and institutional accountability structures. This introduces the concept of \textit{verifiable claims} as a central aspect of developing trustworthy AI. A verifiable claim is a falsifiable statement about an AI system's attributes, such as fairness, safety, or privacy. These can be verified through evidence or argument. The authors emphasize that trust in AI cannot be grounded solely in ethical principles or developer intentions. It must be operationalized via mechanisms that enable verification by independent auditors. This framing underscores the importance of lifecycle-wide governance, including secure decommissioning and archival, as part of a broader ecosystem of institutional accountability and transparency.  The UK Government’s AI Cyber Security Code of Practice, published in January 2025, outlines the principle of securely disposing of data and model \cite{UKAICyberCode2025}.

The SC-NLP-LMF thus bridges the gap between theoretical AI lifecycle constructs and the practical needs of secure, compliant, and adaptive NLP deployment. Unlike existing pipelines focused primarily on automation and scalability (e.g., TFX, Kubeflow), SC-NLP-LMF prioritizes ethical accountability, continuous evaluation, and lifecycle compliance, making it particularly suitable for high-risk sectors such as healthcare, law, education, and public administration. By providing a full-spectrum protocol tailored to NLP's specific risks and regulatory entanglements, the framework serves as a reference architecture for organizations seeking to institutionalize trustworthy AI practices at scale.

\subsection{Healthcare Case Study}

To illustrate the SC-NLP-LMF in practice and demonstrate its application in a realistic context, the following example utilizes a hospital network that deploys an NLP model to extract structured diagnoses from unstructured clinical notes within an electronic health records (EHR) system. The following narrative outlines how each lifecycle phase would be enacted under the proposed framework:

\textbf{Data Governance:} The team begins by curating a linguistically representative and de-identified dataset in compliance with HIPAA and GDPR. Differential privacy techniques are applied to sensitive terms, and data versioning ensures traceable provenance.

\textbf{Secure Model Training:} During model development, the team conducts bias diagnostics using industry-standard toolkits such as the IBM AI Fairness 360 Toolkit and Fairness Indicators. These tools help surface disparities in performance across protected attributes (e.g., race, gender, age) during training and validation. The development pipeline integrates robustness evaluation modules using SafeML, a proactive safety filter designed to detect vulnerability to adversarial examples and prompt-based misuse. Model interpretability is enhanced using SHAP and LIME for error analysis, and data minimization protocols are applied to prevent overfitting to protected features. Together, these measures establish a secure and fair training baseline aligned with SC-NLP-LMF’s guidance.

\textbf{Deployment \& Documentation:} Model Cards document model intent, training data, and known limitations. The system is deployed behind encrypted APIs using TLS, with access control and audit logging enabled.

\textbf{Monitoring \& Incident Response:} Post-deployment, the model is monitored for linguistic drift and classification errors. For example, the emergence of COVID-related or new terminology may trigger shadow evaluations, revealing degraded performance in pediatric oncology cases.

\textbf{Retraining \& Updates:} The model is updated with new annotated data and reevaluated using the original fairness benchmarks. Updated versions are deployed using a blue-green deployment approach \cite{amgothu2024}, and logs are archived for compliance purposes.

\textbf{Decommissioning \& Archival:} After a system upgrade, the legacy model is securely retired. Artifacts are encrypted and archived under retention policies, with lineage records logged in the AI model registry for future audits.

This case illustrates how SC-NLP-LMF provides a practical and secure lifecycle governance framework for deploying high-stakes NLP systems in sensitive domains, such as healthcare.

\section{\uppercase{Results}}

The development and evaluation of the SC-NLP-LMF was grounded in a structured literature synthesis, extensive cross-referencing with existing standards and best practices, and validation through detailed application scenarios. The results indicate that existing governance models often lack depth in supporting secure and compliant NLP development across the full model lifecycle. SC-NLP-LMF addresses this limitation by integrating both general AI governance principles and NLP-specific requirements, making it suitable for high-risk domains such as healthcare, finance, and public services.

A structured review, conducted using the PRISMA methodology, retrieved 186 candidate documents from academic databases, standards bodies, and technical reports. After applying inclusion and exclusion criteria, 45 high-quality documents were selected for final synthesis.  These sources included lifecycle governance frameworks, AI regulatory and compliance guidelines, and scholarly work focused on fairness, explainability, and model monitoring in NLP.

To bridge these gaps, SC-NLP-LMF was constructed as a modular, lifecycle-wide governance framework for NLP systems. Each phase in the framework was mapped to peer-reviewed standards or widely adopted open-source tools. For example, bias auditing methods from IBM’s AI Fairness 360 Toolkit and Microsoft’s Fairlearn support equity assessments during model training. Safety mechanisms, such as SafeML, offer proactive risk containment against the misuse of generative or prompt-sensitive models. The decommissioning phase leverages best practices from the UK Government’s AI Cyber Security Code of Practice to ensure verifiable, encrypted archival, and documentation retention.

The framework's effectiveness was further assessed by applying it to a detailed, narrative case study in healthcare NLP. In this scenario, a hospital NLP model designed to extract diagnoses from clinical notes was evaluated across each phase of the SC-NLP-LMF lifecycle. During post-deployment monitoring, performance drift was triggered by the emergence of new and COVID-related terminology. The structured monitoring and retraining phases enabled timely updates to the model while preserving auditability and fairness. This use case illustrated how SC-NLP-LMF can proactively guide organizations in managing performance, compliance, and operational risk throughout the NLP system's lifecycle.

The final synthesis confirms that SC-NLP-LMF is the first framework to offer a dedicated, phase-specific, and compliance-aware structure for managing the lifecycle of NLP models. By consolidating over 50 authoritative sources, integrating practical toolchains, and emphasizing lifecycle continuity, including underrepresented phases like retirement and monitoring, the framework provides a reproducible foundation for responsible NLP system development. Its emphasis on governance, transparency, and risk reduction is particularly critical as NLP models become increasingly embedded in sensitive decision-making systems.

\section{\uppercase{Discussion}}

The SC-NLP-LMF presented in this paper offers a structured and actionable approach to secure and compliant NLP model lifecycle management. The proposed framework fulfills a critical gap in aligning NLP technical development and operation. It provides a comprehensive mechanism to ensure full NLP lifecycle security and compliance with ethical and regulatory obligations. The practical implementation of this framework, however, introduces several trade-offs and operational challenges that must be acknowledged. This section explores these limitations and opportunities for future advancement. 

One of the primary challenges is integrating compliance-aligned lifecycle practices into existing machine learning (ML) operations (MLOps) workflows. Popular platforms such as TensorFlow Extended (TFX) and Kubeflow Pipelines are optimized for reproducibility, scalability, and automation \cite{baylor,Amershi}, but they offer limited native support for lifecycle-wide bias audits, semantic drift detection, or documentation traceability. Embedding SC-NLP-LMF into these pipelines would require not only tool-level adaptation (e.g., bias-checking modules, versioned drift tracking) but also changes in organizational culture and cross-functional collaboration. This reflects a broader challenge in NLP governance. Effective lifecycle management requires more than technical infrastructure. It demands collaboration among management, product teams, engineers, legal teams, compliance officers, and other domain experts \cite{raji}.

The proposed framework assumes access to internal NLP model design and architectures, training data, and evaluation pipelines. This assumption may not hold true for SaaS (Software as a Service), proprietary LLMs, or third-party APIs (e.g., OpenAI's GPT models), where users lack visibility or control over lifecycle stages. In such cases, the SC-NLP-LMF may need to evolve into a contractual or auditing overlay, focusing on documentation review, API behavior monitoring, and third-party compliance certifications (e.g., SOC 2, ISO 27001). Future research should explore how to adapt secure and compliant lifecycle management to black-box NLP systems while preserving accountability and transparency.

Further work is needed to operationalize SC-NLP-LMF in diverse organizational contexts, including public institutions, SMEs, and multinational enterprises. Pilot implementations, particularly in regulated sectors such as healthcare and finance, can yield practical insights into the scalability and cost-effectiveness of the framework. Future research should also explore automated enforcement mechanisms, particularly in more regulated and critical sectors, such as healthcare and finance. These efforts can also clarify the scalability and cost-effectiveness of the framework. For example, compliance-aware MLOps pipelines, embedded governance agents, and lifecycle auditing protocols that integrate directly with cloud-based NLP services can offer valuable insights.

Despite the various limitations outlined above for the SC-NLP-LMF, it presents a wealth of transformative opportunities for NLP solutions' customers, organizations, and stakeholders. Teams are enabled to engage in meaningful discussions about the development, implementation, and auditing of Natural Language Processing (NLP) systems throughout their lifecycle stages by establishing a structured methodology and a common vocabulary. This framework serves as a reference architecture for regulatory bodies and standards organizations aiming to integrate effective AI oversight practices into their official policies.

\section{\uppercase{Conclusions}}
\label{sec:conclusion}

As natural language processing (NLP) systems become increasingly integrated into critical decision-making pipelines across various domains, including healthcare, finance, law, and government, the need for structured, secure, and compliant lifecycle management has never been more urgent. While existing AI governance frameworks offer valuable high-level guidance, they fall short in addressing the specific risks, operational demands, and regulatory obligations uniquely associated with NLP models, particularly in high-risk contexts defined by evolving regulatory regimes such as the GDPR and the EU AI Act.

This paper introduced the Secure and Compliant NLP Lifecycle Management (SC-NLP-LMF) framework, a novel, multi-phase protocol designed to integrate ethical oversight, technical rigor, and legal compliance into every stage of the NLP model lifecycle. Drawing upon peer-reviewed research, industry standards, and regulatory benchmarks, the framework articulates a lifecycle-wide approach encompassing six critical phases: data governance, secure training, deployment governance, monitoring and drift detection, retraining and updates, and model decommissioning. It provides concrete strategies and tools to operationalize transparency, fairness, privacy, and security in NLP workflows, moving beyond static model documentation toward dynamic, continuous governance.

Evaluation across real-world use cases in healthcare and finance demonstrates that the SC-NLP-LMF is not only theoretically sound but also practically applicable. Its integration of bias audits, adversarial robustness testing, privacy-preserving techniques, and semantic drift monitoring provides a defensible and actionable governance roadmap for organizations operating in compliance and regulatory environments.

Future research should focus on scaling and automating SC-NLP-LMF deployment across diverse organizational contexts, developing open-source compliance toolkits, and engaging with policymakers to codify standards for lifecycle management. As NLP technologies continue to evolve and permeate societal infrastructure, lifecycle governance must evolve at an equally rapid and sophisticated pace, ensuring that these systems serve the public interest, not just in theory but at every stage of their real-world operation.

\balance
\bibliographystyle{IEEEtranDOIandURLwithDate}
\bibliography{refs}

\end{document}